\documentclass[fleqn,usenatbib]{mnras}

\usepackage{newtxtext,newtxmath}

\usepackage[T1]{fontenc}

\DeclareRobustCommand{\VAN}[3]{#2}
\let\VANthebibliography\thebibliography
\def\thebibliography{\DeclareRobustCommand{\VAN}[3]{##3}\VANthebibliography}

\usepackage{hyperref}
\usepackage[inline]{enumitem}
\usepackage[utf8]{inputenc}
\usepackage{acronym}
\usepackage{amsfonts}
\usepackage{amsmath}
\usepackage{booktabs}
\usepackage{color}
\usepackage{enumitem}
\usepackage{fontawesome}
\usepackage{natbib}
\usepackage{pgfplots}
\usepackage{siunitx}
\usepackage{tikz}
\usepackage{rotating}
\usepackage{xspace}
\usepackage{soul}

\usetikzlibrary{arrows.meta}
\usetikzlibrary{shapes.arrows}
\usetikzlibrary{calc}

\newcommand{\Msun}{\textrm{M}_\odot}

\newcommand{\tage}{$t_\text{age}$}

\DeclareMathOperator*{\argmax}{arg\,max}

\newcommand{\code}[1]{\href{#1}{\includegraphics[scale=0.8,trim=0 0.05cm 0 0]{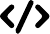}}}

\newacro{BH}[BH]{black hole}
\newacro{GW}[GW]{gravitational wave}
\newacro{PSD}[PSD]{noise power spectral density}

\title[Constraining WDM with simulation-based inference]{Towards constraining warm dark matter with stellar streams through neural simulation-based inference}

\author[J. Hermans et al.]{
Joeri Hermans,$^{1}$\thanks{E-mail: joeri.hermans@doct.uliege.be}
Nilanjan Banik,$^{2}$
Christoph Weniger,$^{3}$
Gianfranco Bertone,$^{3}$
and Gilles Louppe$^{1}$
\\
$^{1}$Montefiore Institute, University of Li{\`e}ge, Belgium\\
$^{2}$Mitchell Institute for Fundamental Physics and Astronomy, Texas A\&M University, College Station, TX 77843, USA\\
$^{3}$GRAPPA Institute, Institute for Theoretical Physics Amsterdam
and Delta Institute for Theoretical Physics, University of Amsterdam, \\
Science Park 904, 1098 XH Amsterdam, The Netherlands
}

\date{Accepted XXX. Received YYY; in original form ZZZ}

\pubyear{2020}

\begin{document}
\label{firstpage}
\pagerange{\pageref{firstpage}--\pageref{lastpage}}
\maketitle

\begin{abstract}
A statistical analysis of the observed perturbations in the density of stellar streams can in principle set stringent contraints on the mass function of dark matter subhaloes, which in turn can be used to constrain the mass of the dark matter particle. However, the likelihood of a stellar density with respect to the stream and subhaloes parameters involves solving an intractable inverse problem which rests on the integration of all possible forward realisations  implicitly defined by the simulation model. 
In order to infer the subhalo abundance, previous analyses have relied on Approximate Bayesian Computation (ABC) together with domain-motivated but handcrafted summary statistics. 
Here, we introduce a likelihood-free Bayesian inference pipeline based on Amortised Approximate Likelihood Ratios (AALR), which automatically learns a mapping between the data and the simulator parameters and obviates the need to handcraft a possibly insufficient summary statistic.
We apply the method to the simplified case where stellar streams are only perturbed by dark matter subhaloes, thus neglecting baryonic substructures, and describe several diagnostics that demonstrate the effectiveness of the new method and the statistical quality of the learned estimator.
\end{abstract}




\section{Introduction}
\label{sec:intro}
Cold Dark Matter (CDM) models~\citep{1982ApJ...263L...1P,1984Natur.311..517B} predict
a hierarchical collapse in which large haloes form through the merging of smaller dark matter clumps~\citep{moore1999dark,avila1998formation,zhao2003growth}.
This process is driven by CDM's scale-free halo mass function~\citep{hofmann2001damping,schneider2013halo} and depends 
on the initial conditions of the matter power spectrum, which in turn 
anticipates the existence of 
dark matter haloes down to $10^{-4}~\Msun$~\citep{bertschinger2006effects}.
Warm Dark Matter (WDM) models~\citep{bond1983collisionless,dodelson1994sterile,2001ApJ...556...93B} on the other hand, in which the dark matter particle is much lighter, influence
structure formation down to the scale of dwarf galaxies. While at large scales the collapse in WDM is hierarchical as well, it becomes strongly suppressed
below the half-mode mass scale of the corresponding dark matter model, where the non-negligible velocity dispersion of dark matter particles prevents haloes to form and smooths the density field instead~\citep{smith2011testing}. Therefore, a powerful method of probing the particle nature of dark matter is to measure the abundances of the lowest mass subhaloes in our galaxy.
While higher mass subhaloes will eventually initiate star formation and manifest themselves as dwarf galaxies, detecting low mass subhaloes ($\lesssim 10^9 ~\Msun$) remains particularly hard since they either have very few faint stars or none at all.

\medskip

Cold stellar streams that formed due to the tidal disruption of globular clusters by the Milky Way potential are a powerful probe for detecting and measuring the abundances of these low mass subhaloes  \citep{ibata2002uncovering,johnston2002lumpy,yoon2011clumpy,carlberg2012dark,Erkal2015,Erkal2015a}. 
When a subhalo flies past a stellar stream, it gravitationally perturbs the orbit of the stream stars around the point of closest approach, which leaves a visible imprint in the form of a region of low stellar density or a \emph{gap}. Such gaps can be individually analyzed to infer the properties of a single subhalo perturber \citep{Erkal2015a}. However, a stream is expected to encounter many subhalo impacts over its dynamical age, leading to complicated density structures that can be hard to separate into individual gaps. Therefore, a more pragmatic approach is to study the full stream density and statistically infer the subhalo abundance within the galactocentric radius of the stream \citep{Bovy2016a}.

Stream-subhalo encounters are described by various quantities such as the impact parameter,
the flyby velocity
of the subhalo, mass and size of the subhalo, and the time and angle of the subhalo impact.
While simulating stream-subhalo encounters and their effects on the stellar density through these parameters is relatively straightforward,
the forward model does not easily lend itself to statistical inference.
The reason for this is that the likelihood of a stellar density with respect to these parameters involves solving an
intractable inverse problem which rests on the integration
of all possible forward realisations 
implicitly defined by the simulation model.
It remains however possible to infer the underlying probabilities by relying on likelihood-free approximations \citep{cranmer2020frontier}.
From this perspective, \citet{Bovy2016a} applied Approximate Bayesian Computation (ABC)~\citep{rubin1984bayesianly} to infer subhalo abundance using the power spectrum and bispectrum of the stream density as a summary statistic. With the same ABC technique, \citet{banik2018probing,banik2019novel} applied the stream density power spectrum as a summary statistic to infer the particle mass of thermal relic dark matter.

\medskip

It should be noted that ABC posteriors are \emph{only} exact whenever the handcrafted summary statistic is \emph{sufficient}, and the distance function chosen to express the similarity between observed and simulated data tends to 0, which in practice is never achievable.
We address this issue by introducing a likelihood-free Bayesian inference pipeline based on amortised approximate likelihood ratios~(\textsc{aalr})~\citep{2019arXiv190304057H}, which automatically learns a mapping between the data and the simulator parameters by solving a tractable minimization problem. 
Afterwards, the learned estimator is
able to accurately approximate the posterior density function of arbitrary stellar streams supported by the simulation model.
By automatically learning this relation from data, we obviate the need to handcraft a possibly insufficient summary statistic,
therefore enabling domain-scientists to pivot from solving the intractable inverse problem to the more natural forward modeling.
In addition, we describe several diagnostics to inspect the statistical quality of the learned estimators with respect to the simulation model.
We demonstrate the effectiveness of this method by inferring the particle mass of dark matter within the stellar stream framework.

\bigskip

The paper is outlined as follows. In Section~\ref{sec:stream_modeling} we present the steps to forward model the stream-subhalo encounter simulations, and highlight our assumptions.
Section~\ref{sec:method} outlines the statistical formalism and the proposed methodology.
Several diagnostics are discussed to probe the statistical quality.
Section~\ref{sec:experiments} evaluates the proposed methodology.
We conclude in Section~\ref{sec:conclusion}. 

\medskip

To support the reproducibility of this work,
we document and provide all code on GitHub\footnote{Available at \href{https://github.com/JoeriHermans/constraining-dark-matter-with-stellar-streams-and-ml}{\texttt{https://git.io/JUvmj}}.}.
A tutorial demonstrating the technique on a toy problem is provided as well.
Steps to obtain the simulated data and pretrained models are described there.
Additionally, we annotate every result and figure with \code{https://github.com/JoeriHermans/constraining-dark-matter-with-stellar-streams-and-ml}, which links to the code or Jupyter notebook used to generate it.

\section{Stream modeling}
\label{sec:stream_modeling}

We use the \texttt{streampepperdf}
simulator\footnote{Available at \url{https://github.com/jobovy/streamgap-pepper}~.} 
that is based
within the \texttt{galpy} framework \citep{bovy2015galpy} to model stream-subhalo interactions. 
Baryonic structures in our galaxy, namely, the bar, spiral arms and the Giant Molecular clouds can induce stream density variations similar to those caused by subhalo impacts \citep{Amorisco2016,Erkal2017,Pearson2017,banik2019effects}. However, owing to its retrograde orbit and a perigalacticon of $\sim 14$ kpc, the effect of the baryonic structures on the GD-1 stream \citep{Grillmair2006} is expected to be subdominant compared to that by a CDM like population of subhalos. Therefore, we have used the GD-1 stream for our analyses and ignored the effects from the baryonic structures. Since the location of GD-1's progenitor is not known, we adopt the model presented in \citet{Webb_2019}, which proposes that the progenitor cluster disrupted in its entirety approximately 500 Myr ago and resulted in the gap at the observed stream coordinate $\phi = -40^{\circ}$. The dynamical age of the GD-1 stream is also unknown and so following the arguments in \citep{banik2019evidence}, we consider all stream models in the range of 3-7 Gyr.

\medskip

Our simulation model samples subhaloes in the sub-dwarf-galaxy mass range $[10^{5} - 10^{9}] \ \Msun$, since density perturbations due to subhaloes less massive than $10^{5} \ \Msun$ are below the level of noise in the current data.
Warm Dark Matter (WDM) is modeled as a thermal relic candidate which is completely described by its particle mass.
The implementation of the subhaloes follows the same procedure as in \citep{Bovy2016a,banik2018probing,banik2019evidence}.

\medskip

We summarize the salient steps of the forward model for completeness. 
The WDM mass function is modeled following \citet{Lovell2013}:
\begin{equation}
\left(\frac{dn}{dM}\right)_{\rm{WDM}} =  \left(1+ \gamma\frac{M_{\rm{hm}}}{M}\right)^{-\beta} \left(\frac{dn}{dM}\right)_{\rm{CDM}},
\label{eq:wdmcdm}
\end{equation}
where $\gamma = 2.7$, $\beta = 0.99$ and $\left(\frac{dn}{dM}\right)_{\rm{CDM}} \propto M^{-1.9}$. Here, $M_{\rm{hm}}$ is the half-mode mass that quantifies the scale below which the mass function is strongly suppressed. Both the CDM and WDM mass functions were obtained by fitting the subhaloes within a Milky Way like analogue from the Aquarius cosmological simulations \citep{Springel2008}. Being dark matter only simulations, these mass functions do not account for the disruption of subhaloes due to baryonic structures, which have been shown to be capable of destroying around $\sim 10 - 50 \%$ of the subhaloes within the galactocentric radius of the GD-1 stream and in the mass range $10^{6.5} - 10^{8.5}~\Msun$ \citep{DOnghia2010,Sawala2016,GarrisonKimmel17a,Webb20b}. 
Moreover, the disrupted fraction of WDM subhaloes is expected to be even higher due to their lower concentrations.
In this paper we ignore subhalo disruptions due to baryonic effects.

\medskip

For each simulated stream density, we consider the region $-34^{\circ} < \phi < 10^{\circ}$ in the observed coordinate frame, and normalize the stream density by dividing it by the mean density. The latter step is different from what was done in \citep{Bovy2016a,banik2019evidence}, where the authors normalize the stream density by dividing it by a $3^{\rm{rd}}$ order polynomial fit.
We tested that both normalization procedures gave similar results. This was also demonstrated in \citet{Bovy2016a}, where they found that changing the order of the smoothing polynomial did not significantly affect the (ABC) posterior. Finally, noise is added to every simulated stream density by sampling a Gaussian realisation of the noise from the observed GD-1 data from \citet{Boer2019}.

\section{Method}
\label{sec:method}
\subsection{Statistical formalism}
\label{sec:method_statistical_formalism}
This work considers two inference scenarios.
In the first we jointly infer the WDM mass $m_\textsc{wdm}$ and the stream age \tage.
The second scenario solely considers $m_\textsc{wdm}$ while marginalizing the stream age.
Because our methodology generalizes to various domains,
we ease the discussion by simplifying the nomenclature into the following concepts:

\bigskip

\noindent{\bf Target parameters $\vartheta$} denote the main parameters of our simulation model.
Depending on the inference scenario at hand, $\vartheta\triangleq(m_\textsc{wdm},~$\tage$)$ or $\vartheta\triangleq(m_\textsc{wdm})$.
Given the Bayesian perspective of this analysis, we define the priors over the WDM mass $m_\textsc{wdm}$ and stream age \tage~to be
$\texttt{uniform}(1,~50)$ keV and $\texttt{uniform}(3,~7)$ billion years (Gyr) respectively.
The upper bound of 50 keV is justified since it corresponds to a half-mode mass of $\sim 4 \times 10^4~\Msun$, which is below the sensitivity of stellar streams given current observational uncertainties.

\bigskip

\noindent{\bf Observables $x$} encapsulate the simulated stellar density of mock streams and the \emph{observed} GD-1 density.
An observable is encoded as a 66-dimensional vector along the linear angle $\phi$ between -34 and 10 degrees.

\bigskip

\noindent{\bf Nominal value $\vartheta^*$} or groundtruth used to simulate the observable $x$ of a mock stream, i.e., $x\sim p(x\vert\vartheta^*)$.

\bigskip

\noindent{\bf Nuisance parameters $\eta$} such as the impact angle and subhalo mass are not of direct interest, but their (random) effects must
be accounted for to infer $\vartheta$ \citep{incidental}. However, this
leaves us with the likelihood function $p(x\vert\vartheta,\eta)$. Given the Bayesian
perspective of this work, we incorporate nuisance parameter uncertainty~\citep{berger1999integrated} by
integration. The priors associated with the nuisance parameters are
implicitly defined through the simulation model.

\subsection{Motivation}
\label{sec:method_motivation}
Our multi-faceted simulation model
induces an extensive space of possible execution paths, which, for example, correspond to randomly sampled dark matter haloes that impact the stellar stream throughout its evolution.
The evaluation of the likelihood $p(x\vert\vartheta)$ of an observable $x$
therefore involves amongst others the integration over a large variety of possible collision histories that are consistent with $\vartheta$.  Given the high-dimensional nature of this integral, directly evaluating data likelihoods is intractable.

\medskip

A common Bayesian approach to address the intractability
of the likelihood is
to reduce the dimensionality of an observable $x$ by means of a
summary statistic $s(x)$.
The reduction in dimensionality
allows the posterior to be approximated numerically
by collecting samples $\vartheta\sim p(\vartheta)$
for which
observables produced by the forward model
$s(x)\sim p(x\vert\vartheta)$ are similar,
in terms of some distance, to the compressed
representation of the observed data $s(x_o)$.
This rejection sampling scheme
is commonly referred to as
\emph{Approximate Bayesian Computation}~\citep{rubin1984bayesianly} (ABC)
and is, as the name indicates, \emph{approximate}.
Although the compression of $x$ into a summary statistic makes
the numerical approximation of the posterior tractable,
it may reduce the statistical power of an analysis because
the selected summary statistic often destroys relevant information.
In fact, ABC is \emph{only} exact whenever the summary statistic
is \emph{sufficient} and the distance
function chosen to express the similarity between
between $s(x)$ and $s(x_o)$ tends to 0.
This is in practice never achievable because for a given simulation budget
\begin{enumerate*}[label=(\roman*)]
  \item a small acceptance threshold severely impacts the
rate at which proposed samples are accepted, affecting
the approximation of the posterior density function, and
  \item the \emph{assumed} sufficiency of the summary statistic is virtually never thoroughly demonstrated in practice.
\end{enumerate*}
Despite these shortcomings, ABC has been fruitfully applied
in cosmology to constrain dark matter models within the context of stellar streams~\citep{banik2018probing, bovy2019constraining, banik2019novel},
and more recently gravitational lensing~\citep{2020MNRAS.491.6077G}.

\medskip

This work tackles the \emph{intractability} of the likelihood from a different perspective. Instead of
manually crafting a summary statistic and a distance function with a specific
acceptance threshold, we propose to
learn an amortised mapping
from target parameters $\vartheta$ and observables $x$ to posterior densities by solving a
\emph{tractable} minimization problem.
The learned mapping has the potential
to increase the statistical power of an analysis since the procedure, in contrast to ABC,
\emph{automatically} attempts to learn an internal sufficient summary statistic of the data.
The automated procedure therefore enables domain-experts
to solely focus on the forward modeling of the phenomena of interest, because the method
does not require any consideration whether synthetic observables are compressible
into low-dimensional summary statistics.
Although the proposed method treats the simulation model as a black box,
we would like to point out that it is possible to improve the efficiency
of the minimization problem, provided that latent information
can be extracted from the simulation model~\citep{brehmer2018constraining,Brehmer:2019jyt,brehmer2020mining},
albeit at some implementation cost.
\begin{figure*}
    \centering
    \begin{tikzpicture}[align=center]
        \draw [rounded corners, thick] (-2,2) rectangle (2,-2);
        \draw [black!100, -Latex] (2 - 0.6667, 2 - 0.6667) -- (2 + 0.6667, 2 - 1 * 0.6667);
        \foreach \in in {2,...,4}
            \foreach \hidden in {1,...,5}
                \draw [black!20] (-2 + 1 * 0.667, 2 - \in * 0.6667) -- (0, 2 - \hidden * 0.6667);
        \foreach \hidden in {1,...,5}
            \draw [black!20] (0, 2 - \hidden * 0.6667) -- (2 - 0.6667, 2 - 0.6667);
        \draw [fill=lightgray!10, draw=gray!100] (-2 + 1 * 0.6667, 2 - 2 * 0.6667) circle [radius=0.25];
        \draw [fill=lightgray!10, draw=gray!90] (-2 + 1 * 0.6667, 2 - 3 * 0.6667) circle [radius=0.25];
        \draw [fill=lightgray!10, draw=gray!90] (-2 + 1 * 0.6667, 2 - 4 * 0.6667) circle [radius=0.25];
        \draw [fill=lightgray!10, draw=gray!90] (0, 2 - 1 * 0.6667) circle [radius=0.25];
        \draw [fill=lightgray!10, draw=gray!90] (0, 2 - 2 * 0.6667) circle [radius=0.25];
        \draw [fill=lightgray!10, draw=gray!90] (0, 2 - 3 * 0.6667) circle [radius=0.25];
        \draw [fill=lightgray!10, draw=gray!90] (0, 2 - 4 * 0.6667) circle [radius=0.25];
        \draw [fill=lightgray!10, draw=gray!90] (0, 2 - 5 * 0.6667) circle [radius=0.25];
        \draw [fill=lightgray!10, draw=gray!90] (2 - 0.6667, 2 - 1 * 0.6667) circle [radius=0.25];
        \node [rounded corners, draw, thick, minimum height=0.9cm, minimum width=2cm, shape=rectangle, anchor=west] (log_ratio) at (2 + 0.6667, 2 - 0.6667) {$\log \hat{r}(x\vert\vartheta)$};
        \node [rounded corners, draw, thick, minimum height=0.9cm, minimum width=2cm, shape=rectangle, anchor=west] (log_prior) at (2 + 0.6667, -2 + 0.6667) {$\log p(\vartheta)$};
        \node [rounded corners, draw, thick, minimum height=0.9cm, minimum width=1.5cm, shape=rectangle, anchor=east, right of=log_ratio, xshift=4cm] (discriminator) {$\hat{d}(\vartheta,x)$};
        \path let \p1 = (log_ratio) in node [draw,fill=white,thick,circle,below of=log_ratio,above of=log_prior] (plus) at (\x1, 0) {+};
        \draw [black!100, -Latex] (log_ratio) -- ++(discriminator) node[midway,fill=white] {$\sigma(\log\hat{r}(x\vert\vartheta))$};
        \draw [black!100, -Latex] (log_ratio) -- ++(plus);
        \draw [black!100, -Latex] (log_prior) -- ++(plus);
        \draw [thick, -Latex] (5.5, -2 + 0.3) -- (10.5, -2 + 0.3) node[anchor=north, yshift=-0.5cm, xshift=-0.25cm] {$m_\textsc{wdm}$};
        \foreach \x in {1,10,20,30,40,50}
            \draw (5.5 + \x / 11, -2 + 0.3 - 0.1) -- (5.5 + \x / 11, -2 + 0.3 + 0.05) node[anchor=north, yshift=-0.15cm] {$\x$};
        \draw [thick, line width=0.75mm] (5.6, -2 + 0.3) .. controls (6, -0) and (6, -1.4) .. (10.05, - 2 + 0.3);
        \path let \p1 = (plus) in node [rounded corners, draw, minimum height=0.9cm, thick, minimum width=2cm, shape=rectangle, anchor=west] (pdf) at (4.5, \y1) {$\log \hat{p}(\vartheta\vert x)$};
        \draw [-Latex] (plus) -- (pdf);
        \path let \p1 = (pdf) in node [fill=none, opacity=1] (corner) at (9.75, \y1) {};
        \draw (pdf) -- (corner.center) node[midway,fill=white] {$\exp(\log\hat{p}(\vartheta\vert x))$};
        \node [opacity=1] (intersect) at (9.75, -2 + 0.2) {};
        \draw [-Latex] (corner.center) -- (intersect);
        \draw (9.75, -2 + 0.3) -- (9.75, -2 - 0.6);
        \draw (9.75, -2 - 0.6) -- (-2 - 2 * 0.6667, -2 - 0.6);
        \path let \p1 = (log_prior) in node [-Latex, fill=white] (log_prior_bottom) at (\x1, -2 - 0.6) {$\vartheta$};
        \draw [-Latex] (log_prior_bottom) -- (log_prior);
        \draw (-2 - 2 * 0.6667, -2 - 0.6) -- (-2 - 2 * 0.6667, 0 - 0.667);
        \draw [-Latex] (-2 - 2 * 0.6667, 0 - 0.667) -- (-2, 0 - 0.6667);
        \node [fill=white] (inputs) at (-2 - 0.6667, 0 - 0.6667) {$\vartheta$};
        \draw [-Latex] (-2 - 3 * 0.6667, 0 + 0.6667) -- (-2, 0.6667);
        \node [fill=white] (outputs) at (-2 - 0.6667, 0 + 0.6667) {$x$};
        \node[inner sep=0pt] (stream) at (-2 - 4.5 * 0.6667, 0.6667) {\includegraphics[width=3.5cm]{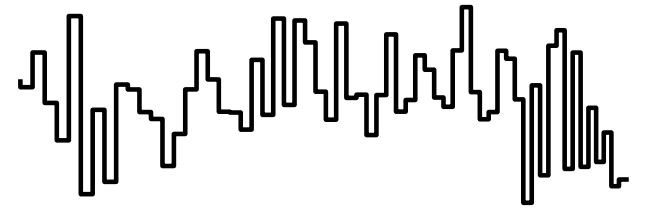}};
        \node (stream_label) at (-2 - 4.5 * 0.6667, 2 * 0.75) {Observed stellar density};
    \end{tikzpicture}
    \caption{Graphical representation of the inference procedure after training the ratio estimator (neural network).
    The ratio estimator accepts a target parameter $\vartheta$ and an observable $x$ as inputs,
    which are subsequently used to approximate the likelihood-to-evidence ratio $\hat{r}(x\vert\vartheta)$.
    The discriminator output $\hat{d}(\vartheta, x)$ --- the sigmoidal projection $\sigma(\cdot)$ of $\log\hat{r}(x\vert\vartheta)$ --- is only used during training. Given that the log prior probability of $\vartheta$ is a tractable quantity, we can easily approximate the log posterior probability $\log\hat{p}(\vartheta\vert x)$ by adding
    the approximated log likelihood-to-evidence ratio. Taking the exponent of the produced quantity results in a direct estimate of the posterior density. This procedure can be repeated for arbitrary target parameters $\vartheta$ supported by the prior.
    It should be noted that the neural network depicted here is an abstract representation. Our technique does not put
    any constraints on the architecture of the neural network. It is therefore possible to use of-the-shelf architectures of arbitrary complexity available in the Machine Learning literature.}
    \label{fig:overview}
\end{figure*}
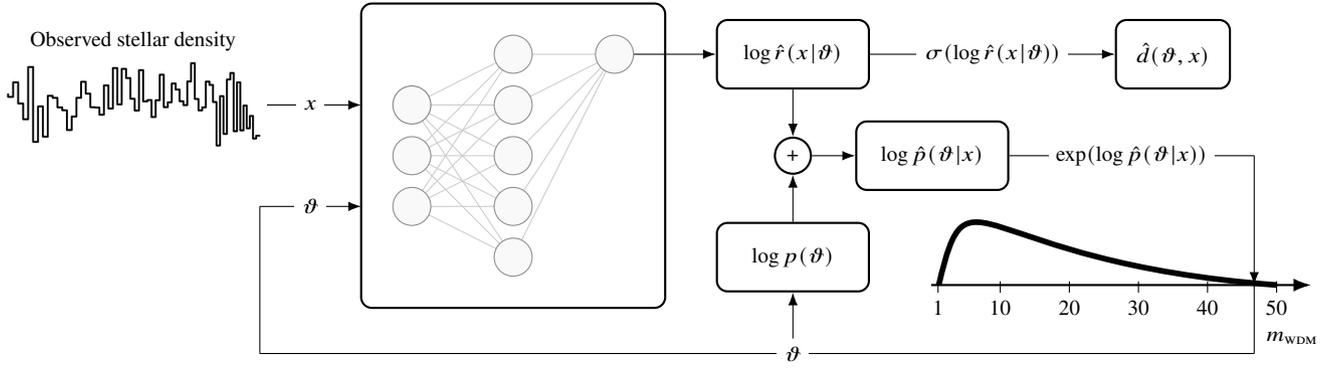
\subsection{Inference}
\label{sec:inference}
The Bayesian paradigm finds model parameters compatible with observation by
computing the \emph{posterior}
\begin{equation}
  \label{eq:bayes}
  p(\vartheta\vert x) = \frac{p(\vartheta)p(x\vert\vartheta)}{p(x)}.
\end{equation}
Evaluating the posterior density for a given target parameter $\vartheta$ and an observable $x$
in our setting is not possible because the
likelihood $p(x\vert\vartheta)$ is per definition intractable.
To enable the tractable evaluation
of the posterior, we have to rely on likelihood-free surrogates
for key components in Bayes' rule.
Note that Equation~\ref{eq:bayes} can be factorized into
the product of the tractable prior probability and the intractable
likelihood-to-evidence ratio $r(x\vert\vartheta)$:
\begin{equation}
    p(\vartheta\vert x) = p(\vartheta)\frac{p(x\vert\vartheta)}{p(x)} = p(\vartheta)\frac{p(\vartheta,x)}{p(\vartheta)p(x)} = p(\vartheta)r(x\vert\vartheta).
\end{equation}
\citet{2019arXiv190304057H} show that an amortised estimator $\hat{r}(x\vert\vartheta)$
of the intractable likelihood-to-evidence ratio can be obtained
by training a discriminator $d(\vartheta, x)$ with inputs $\vartheta$ and $x$, to distinguish between samples from the
joint $p(\vartheta, x)$ with class label 1 and samples from the product of marginals $p(\vartheta)p(x)$ with class label 0 using a discriminative criterion such
as the binary cross entropy.
Whenever the training criterion is minimized,
the authors theoretically
demonstrate that the optimal discriminator $d(\vartheta, x)$
models the Bayes optimal decision function
\begin{equation}
  d(\vartheta, x) = \frac{p(\vartheta, x)}{p(\vartheta,x) + p(\vartheta)p(x)}.
\end{equation}
Subsequently, given a model parameter $\vartheta$ and an observable $x$,
we can use the discriminator as a density \emph{ratio estimator}
to compute the likelihood-to-evidence ratio
\begin{equation}
  r(x\vert\vartheta) = \frac{1 - d(\vartheta,x)}{d(\vartheta,x)} = \frac{p(\vartheta, x)}{p(\vartheta)p(x)} = \frac{p(x\vert\vartheta)}{p(x)} .
\end{equation}
However, the computation of this formulation suffers from significant numerical issues
in the saturating regime where the output of the discriminator tends
to 0. Considering that $\log r(x\vert\vartheta) = \text{logit}(d(\vartheta, x))$ for classifiers
with a \emph{sigmoidal} projection at the output, it is possible to directly obtain
$\log r(x\vert\vartheta)$ from the classifier
by extracting the quantity before the sigmoidal operation.
This strategy ensures that the approximation of $\log r(\vartheta\vert x)$
is numerically stable.
In addition, randomly shuffling $\vartheta$ in a batch
$\vartheta,x\sim p(\vartheta,x)$ instead of drawing a new samples
from the product of marginals significantly aids the convergence rate of the discriminator.
After training, estimates of the posterior probability density
function can be approximated for arbitrary (without retraining)
target parameters $\vartheta$ and observables $x$ by computing
\begin{equation}
  \log p(\vartheta\vert x) \approx \log p(\vartheta) + \log \hat{r}(x\vert\vartheta),
\end{equation}
provided that $\vartheta$ and $x$ are supported
by the prior $p(\vartheta)$ and the marginal model $p(x)$ respectively,
thereby enabling consistent and fast likelihood-free posterior inference.
Figure~\ref{fig:overview} provides a graphical overview.
We refer the reader to \citet{2019arXiv190304057H} or our GitHub repository
for implementation details.

\medskip

The ratio estimator can likewise be adapted to compute a credible region (CR) at a desired level of uncertainty $\alpha$
by constructing a region $\Theta$ in the model parameter space which satisfies
\begin{equation}
  \label{eq:credible_interval}
  \int_\Theta p(\vartheta)r(x\vert\vartheta)~d\vartheta = 1 - \alpha.
\end{equation}
Since many such regions $\Theta$ exist, we select the
highest posterior density region, which is the smallest CR.

\medskip

Although our analysis
focuses on the Bayesian paradigm, it is possible use the ratio estimator in a frequentist setting~\citep{cranmer2015approximating,Brehmer:2019jyt}.
The likelihood-ratio $\lambda(x\vert\vartheta_0,\vartheta_1)$ between two hypotheses $\vartheta_0$ and $\vartheta_1$ can
easily be computed with the ratio estimator
as the denominators of $r(x\vert\vartheta_0)$ and $r(x\vert\vartheta_1)$ cancel out, i.e.,
\begin{equation}
  \label{eq:likelihood_ratio}
  \lambda(x\vert\vartheta_0,\vartheta_1) = \frac{p(x\vert\vartheta_0)}{p(x\vert\vartheta_1)} = \frac{r(x\vert\vartheta_0)}{r(x\vert\vartheta_1)}.
\end{equation}
The same
strategy applies to the likelihood-ratio~\citep{cowan2011asymptotic} test statistic for a specific observable $x$
\begin{equation}
  -2\log\lambda(\vartheta) = -2\log\frac{p(x\vert\vartheta)}{p(x\vert\hat{\vartheta})},
\end{equation}
where the maximum likelihood estimate $\hat{\vartheta}$ is
\begin{equation}
  \hat{\vartheta} = \argmax_{\vartheta} r(x\vert\vartheta).
\end{equation}
The test statistic can thus be expressed~\citep{cranmer2015approximating} as
\begin{equation}
\label{eq:profile_likelihood}
  -2\log\lambda(\vartheta) = -2\left[\log r(x\vert\vartheta) - \log r(x\vert\hat{\vartheta})\right].
\end{equation}
As a result of Wilks' theorem~\citep{wilks1938large}, we can directly convert
the test statistic into a confidence level (CL)
under the assumption that the statistic is $\mathcal{X}_k^2$-distributed
with $k$ degrees of freedom (in function of $\vartheta$'s dimensionality).
\begin{figure*}
    \centering
    \includegraphics[width=\linewidth]{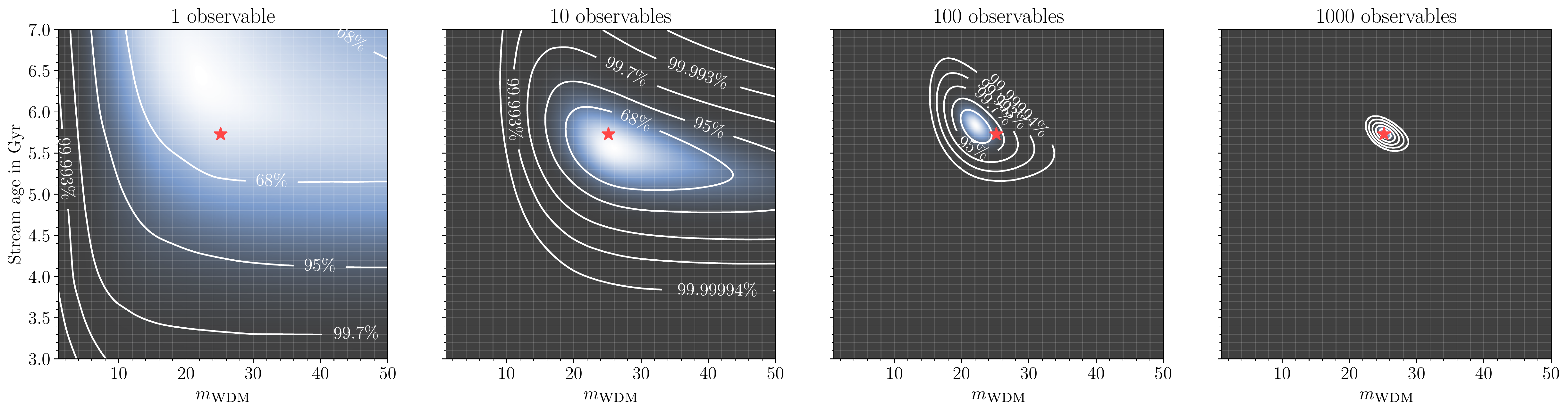}
    \caption{Demonstration of the mode convergence diagnostic described in Section~\ref{sec:diagnostic_mode}.
      The figures show, from left to right, the posteriors for 1, 10, 100 and 1000 independent and identically
      distributed mock GD-1 observables.
      Every figure adopts the same nominal value or groundtruth, which is highlighted by the red star. As the amount of observables increases,
      the posteriors are becoming increasingly more tight around the nominal value.
      This indicates that the individual posteriors do not, in expectation, introduce significant bias for independent and
      identically distributed observables.
      ~~\protect\code{https://github.com/JoeriHermans/constraining-dark-matter-with-stellar-streams-and-ml/blob/master/experiments/experiment-inference/out/diagnostic-map-convergence.ipynb}}
    \label{fig:diagnostic_map_convergence}
\end{figure*}

\subsection{Diagnostics}
\label{sec:diagnostics}
Before making any scientific conclusion, it
is crucial to verify the result of the involved statistical computation.
This is especially challenging in the likelihood-free setting because
evaluating the likelihood is intractable.
The following subsections describe several diagnostics to
assess the quality of the amortised ratio estimates.
No additional training or fine-tuning is applied as this would
change the statistical properties of the ratio estimator.

\subsubsection{Proper probability density}
\label{sec:diagnostic_proper_density}
A ratio estimator $\hat{r}(x\vert\vartheta)$ which correctly
models the true likelihood-to-evidence ratio should satisfy
\begin{equation}
  \label{eq:diagnostic_proper_density}
  \int_\vartheta p(\vartheta)\hat{r}(x\vert\vartheta)~d\vartheta \approx 1~\forall x.
\end{equation}
The diagnostic should be applied to observables $x$
of an evaluation dataset \emph{and} real observables $x_o$. Passing
the diagnostic on the evaluation dataset, while failing
on $x_o$ indicates that $x_o$ is not
supported by the marginal model $p(x)$, because ratio
estimates in this regime are undefined and
can therefore take on arbitrary values.

\subsubsection{Coverage}
\label{sec:diagnostic_coverage}
Coverage quantifies the
reliability of a statistical method to reconstruct the nominal value~\citep{neymanconstruction,schall2012empirical,strege2012fundamental,2013arXiv1301.3166P}.
The approximation of the ratio estimator can thus be assessed by
determining whether the empirical coverage probability matches the
nominal coverage probability, which corresponds to the confidence level $1 - \alpha$.
The empirical coverage probability is estimated using samples from a (large) presimulated evaluation dataset.
This evaluation dataset consists of samples $\vartheta,x\sim p(\vartheta,x)$.
For every pair $(\vartheta, x)$ in the evaluation dataset, we compute
the corresponding credible or confidence interval.
The fraction of samples for which the groundtruth was contained within the interval corresponds to the
empirical coverage probability.
If the empirical coverage probability $\geq 1 - \alpha$, then the ratio estimator passes the diagnostic.
It is of course desirable that the empirical coverage probability of the ratio estimator converges to the confidence level.
A substantially larger empirical coverage probability corresponds to intervals which are overly conservative.
This implies that the ratio estimates are wrong, \emph{but}, that in expectation the estimated posteriors are conservative,
which is not an undesirable property.
It should be noted that coverage can only be computed efficiently
because our ratio estimator amortizes the estimation of the likelihood-to-evidence
ratio. An equivalent study for ABC would have a significant computational cost.

\subsubsection{Convergence of the mode towards the nominal value}
\label{sec:diagnostic_mode}
The diagnostic is based on the idea that the maximum a posteriori (\textsc{map}) estimate
converges towards the nominal value $\vartheta^*$ for an increasing number of independent and
identically distributed observables $x \sim p(x\vert\vartheta^*)$.
If the approximation of $\hat{r}(x\vert\vartheta)$ is correct,
the \textsc{map} should in the limit coincide with the nominal value $\vartheta^*$.
Let $\mathcal{X} = \{x_1,\ldots,x_n\}$ be a set of i.i.d. observables.
To compute the \textsc{map}, we need $p(\vartheta\vert\mathcal{X})$.
As noted by~\citet{Brehmer:2019jyt}, Bayes' rule can be rewritten as
\begin{align}
  p(\vartheta\vert\mathcal{X}) &= \frac{p(\vartheta)\prod_{x\in\mathcal{X}}p(x\vert\vartheta)}{\int p(\vartheta^{'})\prod_{x\in\mathcal{X}}p(x\vert\vartheta^{'})~d\vartheta^{'}}\\
  &= p(\vartheta)\left[\displaystyle\int p(\vartheta^{'})\prod_{x\in\mathcal{X}}\frac{p(x\vert\vartheta^{'})}{p(x\vert\vartheta)}~d\vartheta^{'}\right]^{-1}\\
  &\approx p(\vartheta)\left[\displaystyle\int p(\vartheta^{'})\prod_{x\in\mathcal{X}}\frac{\hat{r}(x\vert\vartheta^{'})}{\hat{r}(x\vert\vartheta)}~d\vartheta^{'}\right]^{-1}.
\end{align}
The integral can be estimated through Monte Carlo sampling.
By checking whether the \textsc{map} concurs with the nominal value, we effectively probe the bias.
Ideally, this diagnostic should be repeated for various groundtruths to
inspect the behaviour of the ratio estimator over the complete model parameter space.
In some settings however, the posterior may be multi-modal.
In such scenarios the convergence of the mode(s) instead of the \textsc{map} should be assessed.
A trial of the diagnostic is shown in Figure~\ref{fig:diagnostic_map_convergence}.

\subsubsection{Receiver operating characteristic}
\label{sec:diagnostic_roc}
We note that $\hat{r}(x\vert\vartheta)$ is only exact whenever
\begin{equation}
  p(x)\frac{p(x\vert\vartheta)}{p(x)} = p(x)\hat{r}(x\vert\vartheta) = p(x\vert\vartheta),
\end{equation}
is satisfied for all $\vartheta$ and $x$.
Although $p(x)$ and $p(x\vert\vartheta)$ cannot be evaluated directly, 
it remains possible to sample from these densities. Given samples from the reweighted
marginal model $p(x)\hat{r}(x\vert\vartheta)$, and from a specific
likelihood $p(x\vert\vartheta)$, the idea is that $\hat{r}(x\vert\vartheta)$
can only be equivalent to $r(x\vert\vartheta)$ whenever a classifier tasked to distinguish
between $p(x)\hat{r}(x\vert\vartheta)$ and $p(x\vert\vartheta)$, cannot extract any
predictive features.
The discriminative performance of this classifier can be assessed
by means of a \emph{Receiver Operating Characteristic} (ROC) curve. A diagonal ROC, which has an \emph{Area Under Curve} (AUC) of 0.5, corresponds to a classifier which is insensitive.
In that case, the ratio estimator passes the diagnostic.
We emphasize that the ratio estimator can incorrectly pass the diagnostic whenever
the classifier is not sufficiently expressive.

\subsubsection{Alternative diagnostics}
\label{sec:alternative_diagnostics}
Our list of diagnostics is not exhaustive.
Some diagnostics are specific to our ratio estimator and
can only be computed efficiently because ratio estimates are amortised.
In fact, the development of diagnostics
for the simulation-based inference literature is an active area of research. For more recent
methodologies we refer the reader to~\citet{sbc} and~\citet{dalmasso2020validation}.

\subsection{Overview of the proposed recipe}
\label{sec:pipeline}
\begin{enumerate}
\item{
  Simulate a train and test dataset by sampling from the joint $p(\vartheta,x)$.
  This is done by drawing samples $\vartheta\sim p(\vartheta)$ and conditioning the simulation
  model on $\vartheta$ to generate observables $x\sim p(x\vert\vartheta)$. These simulations
  can be parallelised arbitrarily because the samples are drawn independently. The
  effective number of simulations depends on the problem at hand. In practice 
  additional simulations were added whenever the ratio estimators did not pass the coverage
  diagnostic, or, if we found over-fitting to be a significant issue during training.
}
\item{
  Train several discriminators $d(\vartheta, x)$ on the previously simulated dataset.
  This has several uses.
  First, the ratio estimators can be ensembled to reduce the variance of the approximation.
  Secondly, as there is only a single true likelihood-to-evidence ratio $r(x\vert\vartheta)$,
  the variability of ratio estimates within the ensemble can be used to quickly assess the convergence.
  A significant deviation in the ratio estimates is indicative of a ill-tuned optimization procedure.
}
\item Probe the trained ratio estimators for flaws with the diagnostics. Afterwards, apply the diagnostic described in Section~\ref{sec:diagnostic_proper_density} to the observable(s) $x_o$.
  
\item Compute the posterior $\hat{p}(\vartheta\vert x_o) = p(\vartheta)\hat{r}(x_o\vert\vartheta)$ and the
  desired credible or confidence intervals.
\end{enumerate}
\begin{table*}
    \centering
    \begin{tabular}{lllllll} \toprule
        \multicolumn{7}{c}{Empirical coverage probability} \\ \toprule
        {Architecture}\hspace{1.5cm} & 68\% CR & 95\% CR & 99.7\% CR\hspace{1.5cm} & 68\% CL & 95\% CL & 99.7\% CL\\ \midrule
        \multicolumn{7}{l}{$\hat{r}(x\vert \vartheta)$~with~$\vartheta\triangleq(m_\textsc{wdm})$} \\ \midrule
        \textsc{mlp}\hspace{0.4cm} & $0.685_{~\pm0.004}$\hspace{0.3cm} & $0.954_{~\pm0.002}$\hspace{0.3cm} & $0.997_{~\pm0.001}$\hspace{0.4cm} & $0.750_{~\pm0.004}$\hspace{0.4cm} & $0.968_{~\pm0.002}$\hspace{0.4cm} & $0.999_{~\pm0.000}$ \\
        \textsc{mlp-bn} & $0.687_{~\pm0.006}$ & $0.951_{~\pm0.002}$ & $0.997_{~\pm0.000}$ & $0.760_{~\pm0.003}$ & $0.970_{~\pm0.002}$ & $0.999_{~\pm0.000}$ \\
        \textsc{resnet-18} & $0.667_{~\pm0.004}$ & $0.943_{~\pm0.002}$ & $0.996_{~\pm0.001}$ & $0.721_{~\pm0.005}$ & $0.960_{~\pm0.002}$ & $0.997_{~\pm0.000}$  \\
        \textsc{resnet-18-bn} & $0.672_{~\pm0.004}$ & $0.945_{~\pm0.001}$ & $0.996_{~\pm0.001}$ & $0.736_{~\pm0.003}$ & $0.961_{~\pm0.002}$ & $0.998_{~\pm0.000}$  \\
        \textsc{resnet-50} & $0.671_{~\pm0.005}$ & $0.947_{~\pm0.003}$ & $0.996_{~\pm0.001}$ & $0.726_{~\pm0.005}$ & $0.963_{~\pm0.000}$ & $0.998_{~\pm0.001}$ \\
        \textsc{resnet-50-bn} & $0.678_{~\pm0.004}$ & $0.949_{~\pm0.004}$ & $0.996_{~\pm0.001}$ & $0.743_{~\pm0.002}$ & $0.966_{~\pm0.001}$ & $0.998_{~\pm0.000}$ \\ \midrule
        \multicolumn{7}{l}{$\hat{r}(x\vert\vartheta)$~with~$\vartheta\triangleq(m_\textsc{wdm},~$\tage$)$} \\ \midrule
        \textsc{mlp} & $0.685_{~\pm0.005}$ & $0.953_{~\pm0.002}$ & $0.998_{~\pm0.000}$ &  $0.752_{~\pm0.003}$ & $0.968_{~\pm0.001}$ & $0.999_{~\pm0.000}$ \\
        \textsc{mlp-bn} & $0.685_{~\pm0.004}$ & $0.952_{~\pm0.003}$ & $0.997_{~\pm0.000}$ & $0.758_{~\pm0.003}$ & $0.970_{~\pm0.002}$ & $0.999_{~\pm0.000}$  \\
        \textsc{resnet-18} & $0.666_{~\pm0.005}$ & $0.945_{~\pm0.002}$ & $0.995_{~\pm0.001}$ & $0.724_{~\pm0.005}$ & $0.961_{~\pm0.002}$ & $0.998_{~\pm0.000}$ \\
        \textsc{resnet-18-bn} & $0.671_{~\pm0.003}$ & $0.945_{~\pm0.003}$ & $0.996_{~\pm0.001}$ & $0.736_{~\pm0.004}$ & $0.961_{~\pm0.002}$ & $0.998_{~\pm0.000}$ \\
        \textsc{resnet-50} & $0.674_{~\pm0.006}$ & $0.944_{~\pm0.002}$ & $0.996_{~\pm0.001}$ & $0.740_{~\pm0.004}$ & $0.970_{~\pm0.002}$ & $0.999_{~\pm0.000}$ \\
        \textsc{resnet-50-bn} & $0.677_{~\pm0.004}$ & $0.947_{~\pm0.003}$ & $0.997_{~\pm0.000}$ & $0.738_{~\pm0.004}$ & $0.970_{~\pm0.002}$ & $0.999_{~\pm0.000}$ \\
        \bottomrule
    \end{tabular}
    \caption{
        Results of the overage diagnostic. Architectures with the \textsc{bn} suffix make use of Batch Normalization.
        For all ratio estimator architectures, we report Bayesian credible regions and frequentist confidence intervals.
        Although credible regions do not necessarily have a frequentist interpretation, they are in fact much closer to the nominal coverage probability compared to the
        confidence intervals. On the contrary, the confidence
        intervals have coverage, but are slightly conservative. 
        Our analyses will therefore focus on constraints based on confidence intervals.
        \protect\code{https://github.com/JoeriHermans/constraining-dark-matter-with-stellar-streams-and-ml/blob/master/experiments/experiment-inference/out/summary-coverage.ipynb}
    }
    \label{table:coverage}
\end{table*}

\section{Experiments and results}
\label{sec:experiments}
We demonstrate the usage of our technique on
various synthetic realisations of GD-1.
Diagnostics are applied to probe the statistical quality of the approximated posteriors
under the specified simulation model. By comparing our technique against ABC,
we highlight the gain in statistical power our technique can bring to the scientific community.
We compute \emph{preliminary} constraints on $m_\textsc{wdm}$
based on observations of GD-1~by \emph{Gaia} proper motions \citep{gaia,2016A&A...595A...1G}
and \emph{Pan-STARRS} photometry. 
It should be noted
these constraints only hold under the assumed
simulation model. An analysis of
(simulation) model misspecification is outside the scope of this work.

\subsection{Setup}
\label{sec:experiments_setup}

\noindent{\bf Simulations}~~We follow the simulation formalism
described in Section~\ref{sec:stream_modeling} and the priors defined in
Section~\ref{sec:method_statistical_formalism}. 10 million
pairs $(\vartheta,x)\sim p(\vartheta,x)$ are drawn from the simulation model for training, and
100,000 for testing. The simulations in the training dataset are reused in our ABC analyses.

\medskip

\noindent{\bf Ratio estimator training}~~All architectures are trained with identical hyperparameter settings.
No exhaustive hyperparameter optimization or architecture-search was conducted.
Options such as weight-decay and batch-normalization (\textsc{bn})~\citep{batchnorm} were evaluated to reduce over-fitting.
All ratio estimators
use \textsc{selu}~\citep{selu} activations and were trained using the \textsc{adamw}~\citep{adamw} optimizer for 50 epochs
with a batch-size of 4096. We found that larger batch-sizes, for our setting, generalized better.
Appendix~\ref{appendix:sec:batch_size} investigates the influence of the batch-size on the approximations in detail.
We empirically found \textsc{selu} and \textsc{elu} activations to be preferable over \textsc{relu}-like activations,
because the approximation of the posterior density function was generally smoother.
This work considers 3 main architectures;
\begin{enumerate*}[label=(\roman*)]
    \item a simple feedforward \textsc{mlp},
    and variants to \textsc{resnet}~\citep{resnet} such as
    \item \textsc{resnet-18} and \item \textsc{resnet-50}
\end{enumerate*}. Both use 1 dimensional convolutions without dilations
since the usage of dilated convolutions did not yield any significant improvements in terms of test loss.
Because our methodology treats $\vartheta$ as an input feature, we cannot easily
condition the convolutional layers of the \textsc{resnet}-based architectures on $\vartheta$.
This would require conditional convolutions~\citep{yang2019condconv} or hypernetworks~\citep{ha2016hypernetworks} to generate specialized kernels for a given $\vartheta$.
To retain the simplicity of our architecture, we inject the dependency on
$\vartheta$ in the fully connected trunk
of the convolutional ratio estimators.
Other architectural considerations were not explored.
Appendix~\ref{appendix:sec:appendix_architecture}
lists the hyperparameter settings.

\bigskip

\noindent{\bf Approximate Bayesian Computation}~~
Instead of using the stream density power as summary statistics as in \citet{Bovy2016a,banik2019evidence}, we construct a summary statistics based on the stream density itself. We divide the synthetic observable $x$ (with $n = 66$ bins)
by the observable of interest $x_o$
to obtain the
bin-wise stellar density ratio $d = x / x_o$.
Our summary statistic and distance function are jointly expressed as
\begin{equation}
    s(x) = \frac{1}{n}\sum_{i=1}^{n} (d_i - \bar{d})^2,
\end{equation}
where $\bar{d}$ is the mean stellar density ratio.
Ideally, if both observables match perfectly, then $s(x) = 0$.
The acceptance threshold is tuned such that for any given observable of interest $x_o$,
the number of accepted posterior samples is 0.1\% of the simulation budget,
therefore yielding the smallest threshold with respect to the specified acceptance rate.
This corresponds to approximately 10,000 posterior samples.
Our goal is to 
highlight generic aspects of ABC with respect to the proposed method in terms of tuning of the analyses,
and its statistical quality \emph{for the given simulation budget}.
We emphasize that several scheduling and threshold strategies for ABC exist in the literature, see e.g.~\citep{Lintusaari2017-zn, Prangle2017-pe}. We opted here for a method that is based on the same number of simulations used for training the neural network.  The threshold was chosen heuristically to obtain sufficiently smooth posteriors across the entire parameter space, and was not tuned depending on the WDM mass and stream age.  This is different from the targeted convergence check and simulation strategy in previous ABC-based streams analyses~\citep{Bovy2016a,banik2018probing, bovy2019constraining, banik2019novel}.  We cannot exclude that the ABC results shown here could further improve by significantly increasing the number of simulations beyond what was needed for the neural network training.  This is beyond the scope of the current work.

\subsection{Statistical quality}
\label{sec:statistical_query}
We now assess the statistical properties of the trained ratio estimators.
For every architecture, we select the weights
which achieved the smallest test-loss.

\medskip

\begin{figure}
    \centering
    \includegraphics[width=\linewidth]{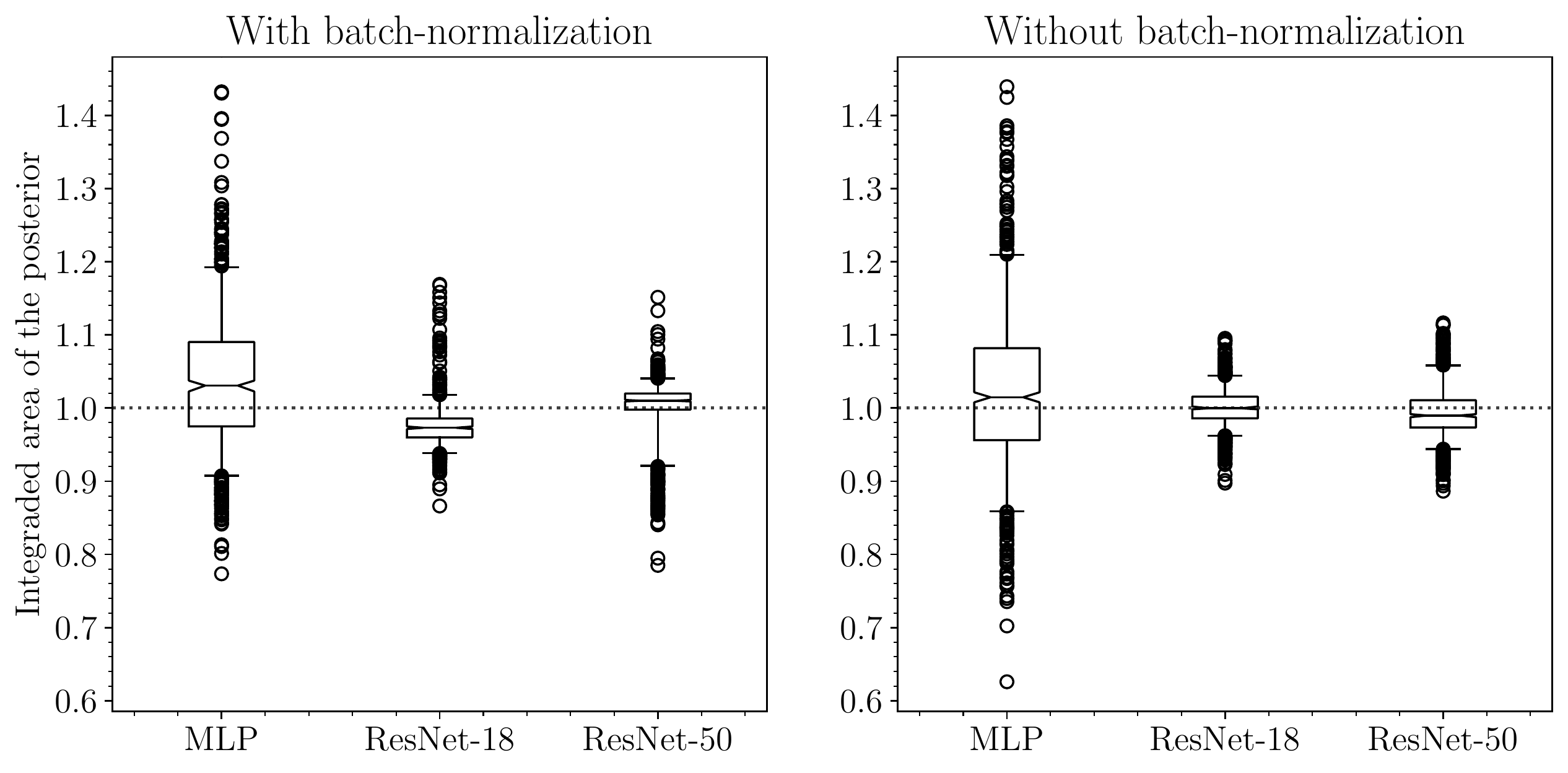}
    \caption{Result of the proper probability density diagnostic.
    As expected, high-capacity models (\textsc{resnet}) have tighter approximations compared to the \textsc{mlp} architectures.
    An interesting discrepancy between the usage of with and without batch normalization is observed.
    \emph{(Left)} With batch-normalization.
    \emph{(Right)} Without batch-normalization.
    ~~\protect\code{https://github.com/JoeriHermans/constraining-dark-matter-with-stellar-streams-and-ml/blob/master/experiments/experiment-inference/out/summary-integrand.ipynb}}
    \label{fig:integrand_boxplots}
\end{figure}
\noindent{\bf Proper probability density}~~The computational
cost of the integration does not allow us to do an exhaustive
analysis. Instead, we apply the diagnostic to 1000 randomly
sampled observables. As before, we repeat the experiment 10 times.
The following results were obtained:
\textsc{mlp} ($1.023\pm0.11$),
\textsc{mlp-bn} ($1.037\pm0.09$),
\textsc{resnet-18} ($1.00\pm0.02$),
\textsc{resnet-18-bn} ($0.973\pm0.03$),
\textsc{resnet-50} ($0.993\pm0.03$),
and \textsc{resnet-50-bn} ($1.001\pm0.04$)
~\protect\code{https://github.com/JoeriHermans/constraining-dark-matter-with-stellar-streams-and-ml/blob/master/experiments/experiment-inference/out/summary-integrand.ipynb}.
Although
the average integrated area under the approximated posterior density
functions approaches 1 for all ratio estimator architectures,
the results suggest that the approximations of the
\textsc{resnet}-based architectures are more robust.
A more careful analysis of the integrated areas, presented in Figure~\ref{fig:integrand_boxplots}, confirms this.
Interestingly, the integrated areas for \textsc{resnet} architectures \emph{with} batch-normalization have a larger
spread compared to their counterparts without batch-normalization.
Our evaluations on GD-1~will therefore focus on
\textsc{resnet}-based architectures without batch-normalization.

\medskip

\noindent{\bf Coverage}~~Table~\ref{table:coverage}
summarizes the empirical coverage probability
of the ratio estimators. For every ratio estimator,
we compute the credible and confidence intervals as described in Section~\ref{sec:inference}.
For both paradigms, we evaluate the interval construction on 10,000 observables,
which is repeated 10 times.
The empirical coverage probability of a ratio estimator is therefore based
on approximately 100,000 observables in total.
We empirically find that \textsc{mlp}-based architectures have
coverage under both Bayesian credible and frequentist
confidence intervals. This is not the case for \textsc{resnet}-based
architectures.
It is noteworthy that the empirical coverage probability
of the credible regions are much closer to the nominal coverage
probabilities
compared to their frequentist counterparts.
Additional statistical power could therefore be extracted if
the credible regions could be tuned to
sufficiently cover the groundtruth at a given nominal coverage probability.
We discuss such an approach based on Neyman construction in Appendix~\ref{appendix:sec:cr_bias}.

\medskip

\begin{figure}
  \centering
  \includegraphics[width=\linewidth]{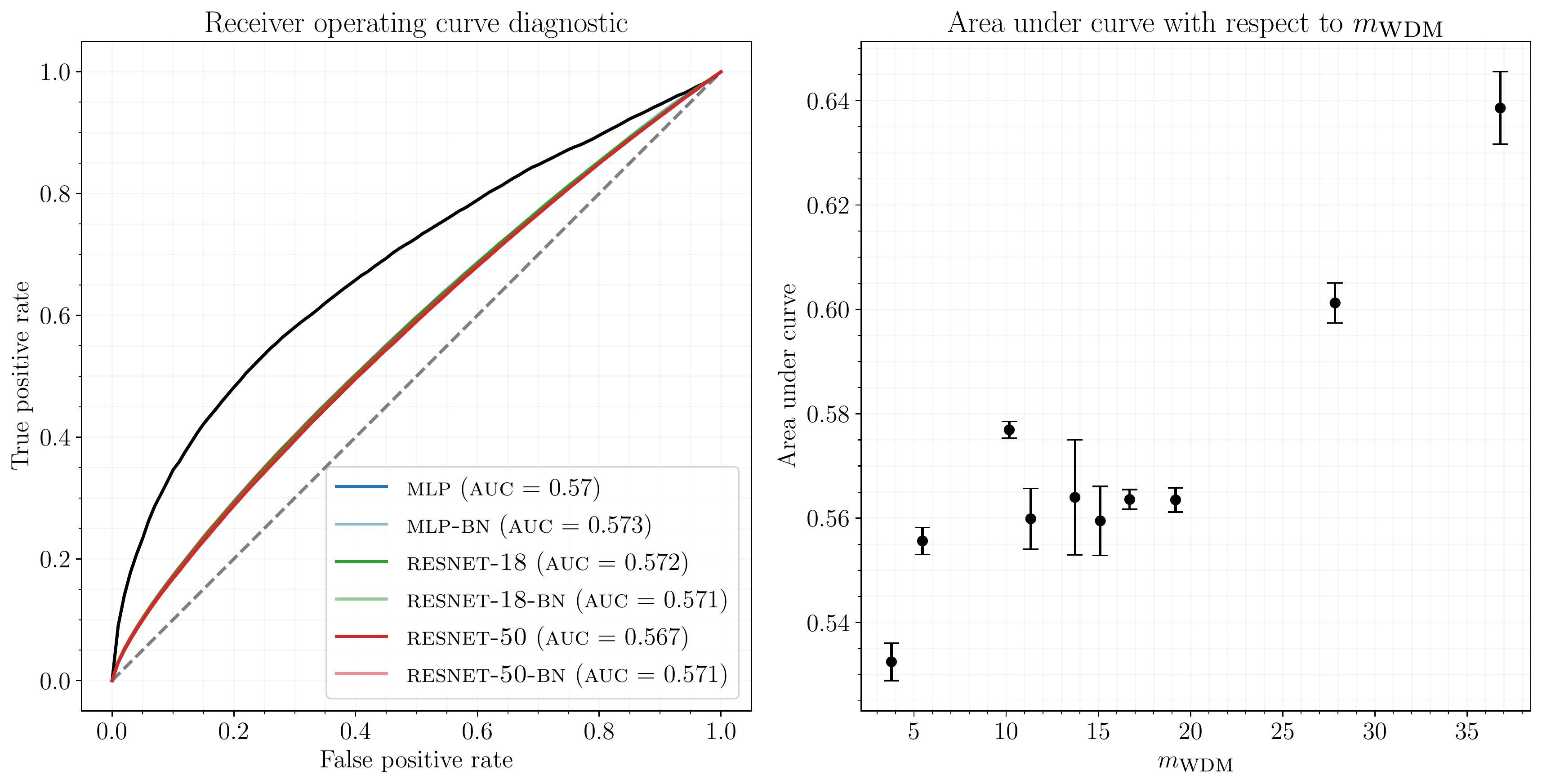}
  \caption{Summary of the receiver operating curve diagnostic.
\emph{(Left)} Area Under Curve (AUC) for all test-hypotheses. A baseline measurement, indicated by the black line, does
not reweigh the marginal model. Although the ratio estimators perform significantly better compared to the baseline, the diagnostic indicates that all ratio estimators do not perfectly approximate the
likelihood-to-evidence ratio (since AUC $\neq0.5$).
This is not necessarily an issue, because the coverage diagnostic demonstrates that the
confidence intervals are conservative. \emph{(Right)} Average AUC of the test-hypotheses under consideration. Larger values of $m_\textsc{wdm}$ are associated with a degraded quality of the ratio estimates.
    ~~\protect\code{https://github.com/JoeriHermans/constraining-dark-matter-with-stellar-streams-and-ml/blob/master/experiments/experiment-inference/out/summary-roc.ipynb}}
  \label{fig:roc_diagnostic}
\end{figure}

\noindent{\bf Receiver operating characteristic}~~We now directly probe
the correctness of the approximated likelihood-to-evidence ratios.
Every ratio estimator is evaluated on 10 uniformly sampled test-hypotheses.
10,000 observables are drawn from every test-hypothesis. For
every test-hypothesis, we repeat the computation of the area
under curve 10 times to account for the stochastic
training of the classifier tasked to distinguish
between samples from the reweighted marginal model
and samples from the test-hypothesis.
Figure~\ref{fig:roc_diagnostic} summarizes the results.
In general, we find that all ratio estimators are unable to perfectly approximate
$r(x\vert\vartheta)$.
This result is not unexpected, because the coverage diagnostic indicates
that the confidence intervals are conservative, which implies that our
estimates of the \emph{true} likelihood-to-evidence ratio must be wrong.
Incorrect, but conservative estimates of the posterior are not a significant issue
because we mainly seek to constrain $m_\textsc{wdm}$.

\medskip

We additionally find that the quality of the ratio estimates degrades
for larger values of $m_\textsc{wdm}$ across all architectures.
Several strategies could be applied to address this. First, 
more expressive architectures could be explored
which potentially make more efficient use of the available data. 
Second, by using additional observables
could be simulated
to aid the approximation of the underlying densities.
In our specific case, a straightforward application of this strategy would be to simulate additional
observables for $\vartheta \gtrapprox 20$ keV. We would like to emphasize
that increasing the size of the training dataset by
simulating additional observables at specific target parameters $\vartheta$ \emph{should not be done},
because this implicitly changes the prior and therefore the underlying marginal model.
Instead, additional observables should only be simulated by sampling from the joint $p(\vartheta, x)$.

\subsection{Evaluation}
\label{sec:evaluation}
\begin{figure}
    \centering
    \includegraphics[width=\linewidth]{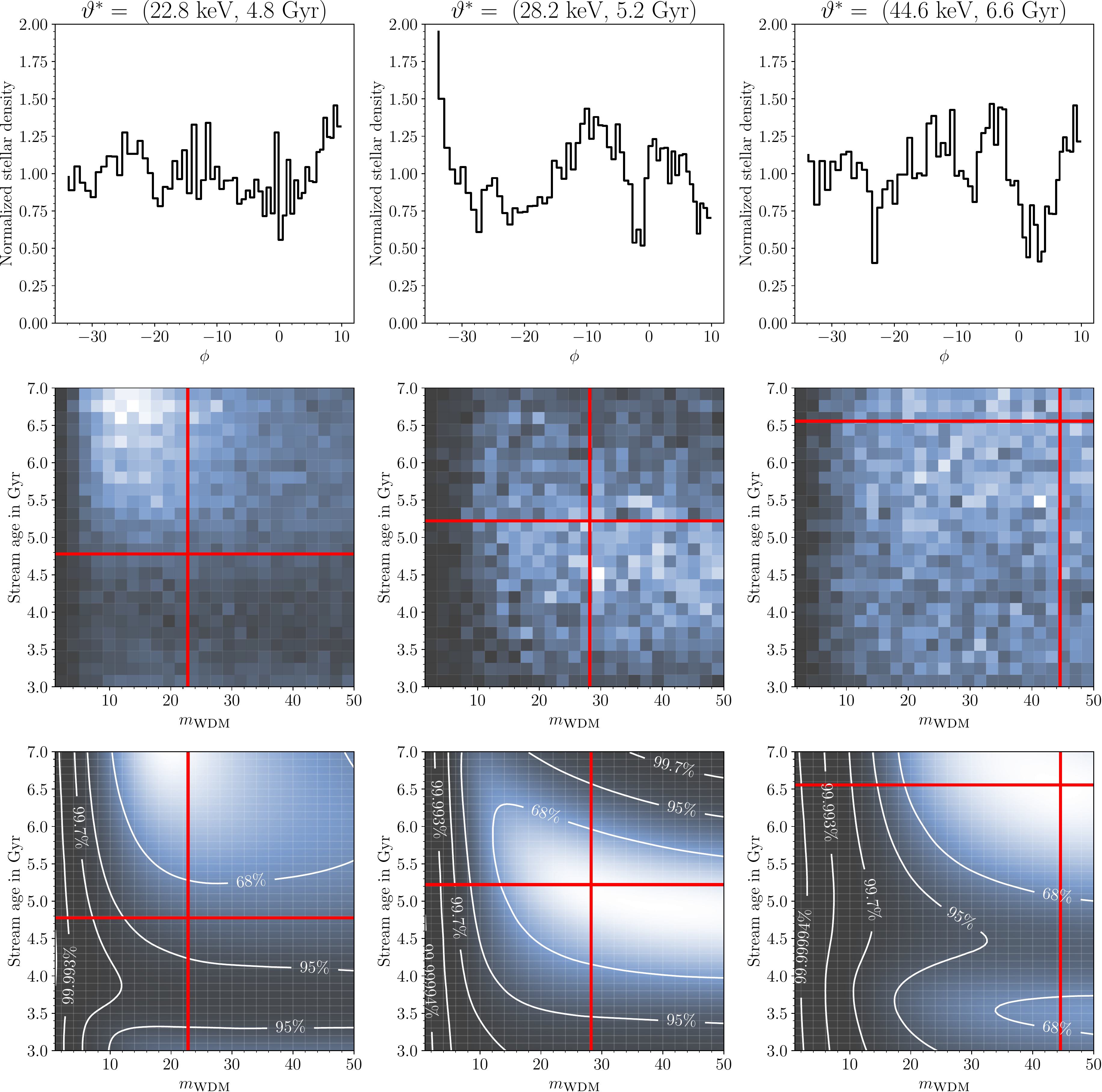}
    \caption{Compact summary of comparisons against ABC.
    All comparisons are listed in Appendix~\ref{appendix:sec:comparison_abc}.
    Every column relates to a single mock simulation. The rows show, from top to bottom,
    the observable, the approximate posterior ABC, and our method respectively.
    The red cross indicates the groundtruth.
    ABC and our method are in agreement for most mock simulations.
    ~~\protect\code{https://github.com/JoeriHermans/constraining-dark-matter-with-stellar-streams-and-ml/blob/master/experiments/experiment-inference/out/summary-abc-new.ipynb}}
    \label{fig:abc_results_compact}
\end{figure}
\begin{figure*}
    \centering
    \includegraphics[width=\linewidth]{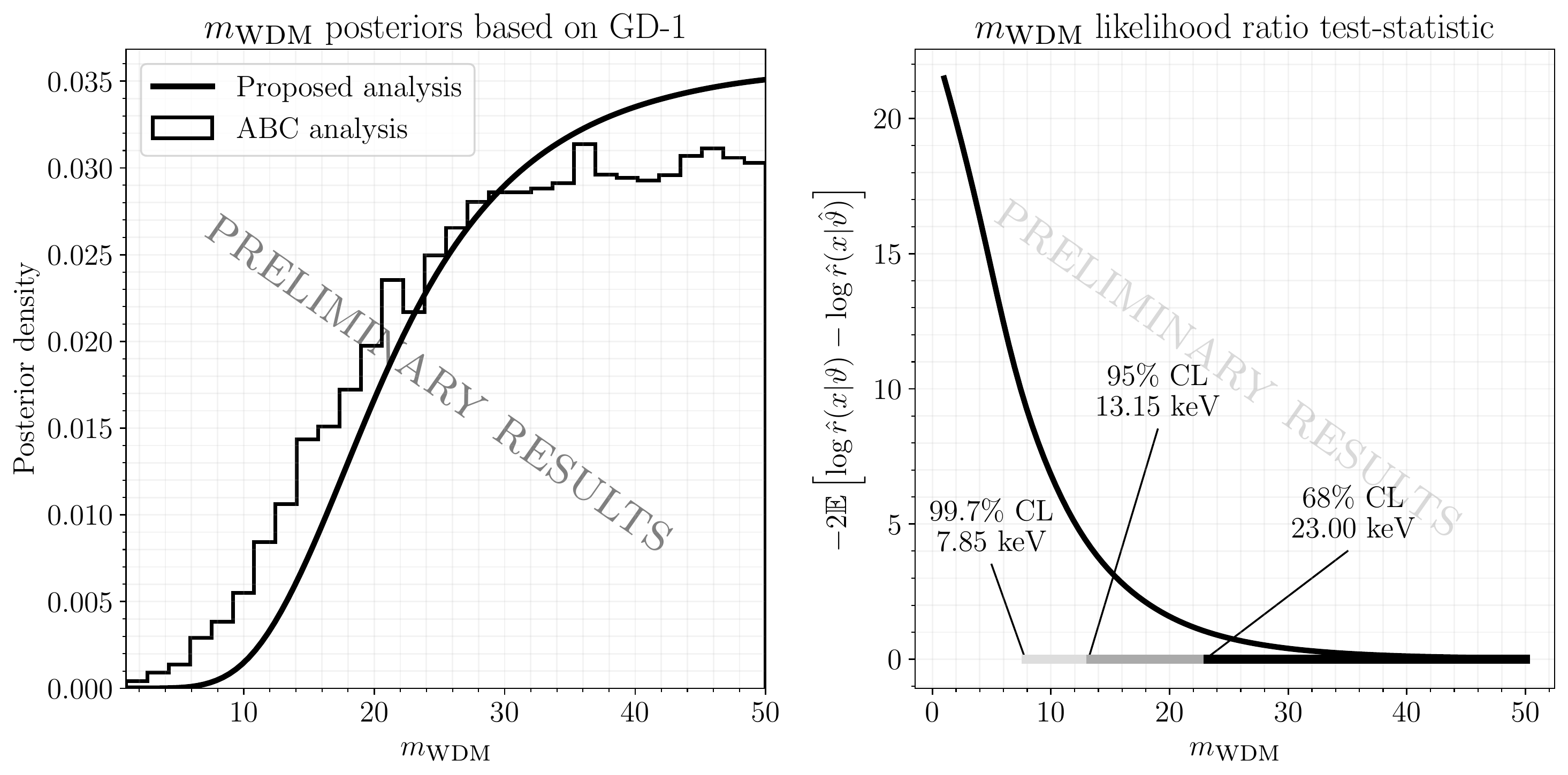}
    \caption{Age-marginalized results based on the observed stellar density variations of GD-1.
    \emph{The results shown here illustrate the power of the proposed methodology, but should be considered as preliminary, since e.g.~baryonic effects are not yet fully included in the simulation model.}
    \emph{(Left)} Direct comparison of the reference ABC and the proposed analysis. Both posteriors indicate a preference for CDM over WDM within
    the assumed simulation model. We find that the proposed method is able to put stronger constraints on $m_\textsc{wdm}$. 
    \emph{(Right)} Likelihood ratio test-statistic used to derive the lower limit confidence intervals.
    ~~\protect\code{https://github.com/JoeriHermans/constraining-dark-matter-with-stellar-streams-and-ml/blob/master/experiments/experiment-inference/out/summary-gd1-inference.ipynb}}
    \label{fig:gd1_posteriors}
\end{figure*}
The performance of both methods is assessed on various randomly sampled GD-1 mock simulations
with distinct nominal target parameters.
A compact overview of the computed posteriors is shown in Figure~\ref{fig:abc_results_compact}. A full overview can be found in Appendix~\ref{appendix:sec:comparison_abc}.
We find the proposed methodology to be preferable over the ABC analysis regarding the reconstruction of the nominal target parameters, and with respect to our stronger, statistically tested, confidence intervals.

In conjunction with the foregoing statistical validation of the ratio estimators, 
these results highlight the fact that ABC requires a carefully crafted summary statistic;
a problem that is absent, or effectively automatised, in the proposed method.
As mentioned earlier, an ABC posterior is only exact whenever the summary statistic is sufficient
\emph{and} the acceptance threshold tends to 0. If these conditions are not met,
the posterior is possibly inaccurate or biased.
The necessity of a sufficient summary statistic
underlines an important issue with ABC in practice;
the \emph{assumed} sufficiency. Determining the statistical
validity of an ABC analysis is computationally demanding and often not feasible.
Our method does not suffer from this issue, because the estimation of the
posterior density is amortised.

\subsection{Towards constraining $m_\text{WDM}$ with GD-1}
\label{sec:experiments_gd1}
We now apply our methodology to obtain a \emph{preliminary} constraint on $m_\textsc{wdm}$,
based on the observed stellar density along the GD-1 stream.
The posteriors in this section are computed using the previously trained and statistically validated
\textsc{resnet}-50 ratio estimator. We would like to remind the reader that the coverage diagnostic indicates
that the derived confidence intervals are slightly conservative.
Our results suggest a strong preference for CDM over WDM.
The posteriors and credible intervals at various confidence levels are shown in Figure~\ref{fig:gd1_posteriors}. 
We find the integrated area under the approximated posterior to be ($0.96 \pm0.011$~\protect\code{https://github.com/JoeriHermans/constraining-dark-matter-with-stellar-streams-and-ml/blob/master/experiments/experiment-inference/out/summary-gd1-inference.ipynb}).
After marginalizing the stream age, the proposed methodology yields $m_\textsc{wdm}\geq 17.5$ keV (95\% CR) and $m_\textsc{wdm}\geq 10.5$ keV (99.7\% CR). 
No significant constraints can be put on the age of GD-1, although an older stream is preferred.
A frequentist perspective based on likelihood ratio limits finds
$m_\textsc{wdm}\geq 13.15$ keV (95\% CL) and $m_\textsc{wdm}\geq 7.85$ keV (99.7\% CL) after marginalizing the stream age. 
Assuming the posterior approximated by ABC is exact, we find 
$m_\textsc{wdm}\geq 10.8$ keV (95\% CL) and $m_\textsc{wdm}\geq 3.5$ keV (99.7\% CL).
We emphasize that \textit{our simulation model does not account for baryonic effects, disturbances caused by massive ($\gtrsim 10^9~\Msun$) subhaloes, and effects induced by variations in the Milky Way potential.}

\medskip

However, our results are promising.  We expect that the proposed method will enable an optimal discrimination between dark matter and baryonic effects (provided the latter can be convincingly modeled). It thus constitutes a powerful probe for constraining the mass of thermal or sterile neutrino dark matter~\citep{dodelson1994sterile,shi1999new,abazajian2001sterile,asaka2005numsm,boyarsky2009role} (although a discrimination between such WDM models might be challenging).

\section{Summary and discussion}
\label{sec:conclusion}
This work proposes a general recipe for the usage of
neural simulation-based inference in the natural sciences.
Although the procedure generalizes to many domains,
we apply our methodology in the stellar stream framework
to determine the nature of the dark matter particle.
We summarize our findings as follows:

\begin{itemize}
    \item Bayesian inference based on Amortised Approximate Likelihood Ratios (\textsc{aalr}) is a powerful and convenient analysis framework to study the statistical properties of density variations of stellar streams. In Figure~\ref{fig:abc_results_compact} we demonstrate that (at least in the absence of the uncertainties from the baryonic effects), GD-1-like streams could be used to simultaneously constrain the mass of thermal relic dark matter and the age of the stream. 
    
    \item \textsc{aalr}, in contrast to ABC, does not require handcrafted summary statistics and tuned acceptance thresholds. Our out-of-the-box \textsc{aalr} analysis are expected to be at least as good as any ABC implementation, and to often significantly outperform ABC, as evident in Figure~\ref{fig:gd1_posteriors}. 
    \item The amortised posterior estimation in \textsc{aalr} allows for a variety of diagnostics, including coverage and bias tests, which are computationally demanding and often infeasible for ABC. We explicitly demonstrate that posteriors approximated by \textsc{aalr} are unbiased and that the corresponding confidence intervals have coverage, as show in in see Figure~\ref{fig:diagnostic_map_convergence} and Table~\ref{table:coverage} respectively.
\end{itemize}

\bigskip

Finally, our preliminary results for GD-1 are promising as they indicate that \textsc{aalr} is an excellent and versatile method to probe the nature of dark matter with stellar streams. At face value, we can probe WDM masses up to 17.5 keV (95\% credible lower limit for a GD-1-like stream).  \emph{We emphasize however that our simulation codes do not account for baryonic effects, which are expected to significantly impact the results.}  In upcoming analyses we plan to use the improved statistical power achieved through \textsc{aalr} to obtain more statistically robust and tighter constraints on the particle mass of dark matter.

\section*{Acknowledgements}
Joeri Hermans would like to thank the National Fund for Scientific Research for his FRIA scholarship.
Gilles Louppe is recipient of the ULiège - NRB Chair on Big data and is thankful for the support of NRB.
All authors would like to thank the developers of the \emph{Jupyter}~\citep{jupyter} project, \emph{Matplotlib}~\citep{matplotlib},
\emph{Numpy}~\citep{numpy}, \emph{PyTorch}~\citep{paszke2017automatic},
and \emph{Scikit-Learn}~\citep{scikit-learn} for enabling this work.

\section*{Data Availability}
The data underlying this article are available in the article and in its online supplementary material.


\bibliographystyle{mnras}
\bibliography{main} 



\clearpage
\appendix
\section{Influence of the batch-size during training on the approximated posteriors}
\label{appendix:sec:batch_size}

\begin{figure}
  \centering
  \includegraphics[width=\linewidth]{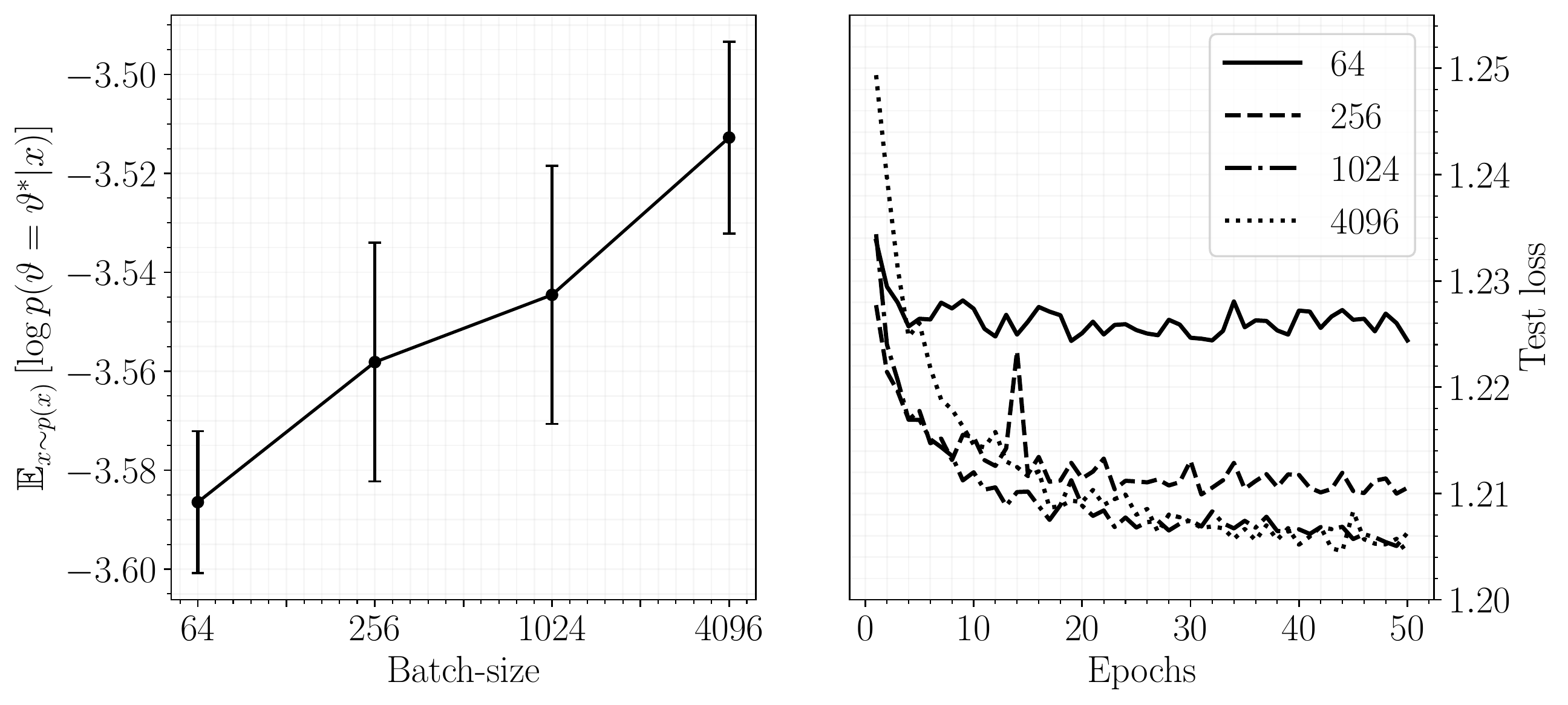}
  \caption{\emph{(Left)} The expected log posterior probability of the nominal $\vartheta^*$ for $\vartheta^*,x\sim p(\vartheta,x)$. Under the assumption that $p(\vartheta)\hat{r}(x\vert\vartheta)$
  is a proper probability density, this quantity captures the ability of the approximated
  posteriors to reconstruct the nominal value.
  \emph{(Right)} Average test-loss curve for various batch-sizes. For a given learning rate,
  there is a clear inverse relation between the test-loss and expected log posterior probability of the nominal value. Larger batch-sizes should therefore be preferred.
    ~~\protect\code{
https://github.com/JoeriHermans/constraining-dark-matter-with-stellar-streams-and-ml/blob/master/experiments/experiment-inference/out/summary-batch-size.ipynb\#Verify-95\%-CR-coverage}}
  \label{fig:batch_size_expectation_loss}
\end{figure}

To investigate the effect of the batch-size on the approximated posteriors,
we train several ratio estimators based on the \textsc{mlp-bn} (batch normalization) architecture with
batch-sizes 64, 256, 1024 and 4096. At 95\% CR, the
empirical coverage probabilities of these ratio estimators are $0.961\pm0.004$, $0.954\pm0.004$, $0.952\pm0.008$ and $0.952\pm0.006$ respectively~\protect\code{https://github.com/JoeriHermans/constraining-dark-matter-with-stellar-streams-and-ml/blob/master/experiments/experiment-inference/out/summary-batch-size.ipynb}.
Figure~\ref{fig:batch_size_expectation_loss} shows the test-loss curves and
$
\label{eq:reconstruct_nominal}
\mathbb{E}_{\vartheta,x\sim p(\vartheta,x)}\left[\log \hat{p}(\vartheta=\vartheta \vert x)\right]
$
for every batch-size setting. Under the assumption that $p(\vartheta)\hat{r}(x\vert\vartheta)$ is
a proper probability density, Equation~\ref{eq:reconstruct_nominal} captures the ability of $\hat{r}(x\vert\vartheta)$
to reconstruct the groundtruth. As indicated by Figure~\ref{fig:batch_size_expectation_loss},
there is a clear negative correlation between the test-loss and the expected log posterior
probability of the nominal value.
Although not entirely unexpected, this suggests that larger batch-sizes
have the potential to further reduce the test-loss at a given learning rate.
Practitioners should therefore analyze the behaviour of their optimization
procedure with respect to the batch-size as well.
\newline

The observations made here are in line with
the machine learning literature~\citep{hoffer2017train},
although others~\citep{keskar2016large,masters2018revisiting} have suggested that smaller batch-sizes
lead to models which generalize to a greater degree. This especially seems to be the case
whenever the testing loss surface differs from the training loss surface~\citep{keskar2016large}.
Unlike typical deep learning applications with a fixed dataset,
this issue can easily be addressed within the likelihood-free setting, because
the similarity of these loss surfaces can be ensured by continuously drawing new samples from
the simulation model. For a given learning rate, larger batch-sizes should therefore be preferred~\citep{smith2017don}.
Alternatively, this could also be explained due to the fact that larger batch-sizes provide more empirical evidence (less stochasticity)
to approximate the ratio.
\section{Hyperparameters}
\label{appendix:sec:appendix_architecture}
The same hyperparameters are used across all architectures.
We did not explore specific settings for every individual architecture,
demonstrating the robustness of our technique.
A learning rate of 0.0001 with a batch-size of 4096 and a weight-decay factor of 0.1 was used during training.
The ratio estimators do not use dropout~\citep{hinton2012improving}.
The remaining hyperparameters (e.g., of Batch Normalization) were set to the \emph{PyTorch} defaults. \protect\code{https://github.com/JoeriHermans/constraining-dark-matter-with-stellar-streams-and-ml/blob/master/experiments/experiment-inference/pipeline.sh\#L34}.

\section{Neyman construction with ratio estimators and Bayesian credible regions}
\label{appendix:sec:cr_bias}
As indicated by Table~\ref{table:coverage}, the method responsible for computing the Bayesian credible regions
closely approximates the nominal coverage probability, even though credible regions
do not necessarily have a frequentist interpretation. For most ratio estimators however,
the method does not sufficiently cover the groundtruth.
The credible regions in question are derived from the intersection
between the posterior density and the highest density
such that the area below the intersection is approximately $1 - \alpha$.
Credible regions can therefore be made more conservative
by artificially lowering the highest density level until they have
coverage at some given confidence level.
We achieve this by introducing a bias term $\alpha_b$ such that
the integrated area under the credible region $\Theta$ is $1 - \alpha - \alpha_b$.
Using the same ratio estimators as in Table~\ref{table:coverage}, we repeat the
experiment with $\alpha_b$ 0.002, 0.02 and 0.02 across all architectures for 68\% CR, 95\% CR and 97.7\%\ CR
respectively~\protect\code{https://github.com/JoeriHermans/constraining-dark-matter-with-stellar-streams-and-ml/blob/master/experiments/experiment-inference/pipeline.sh\#L343}.
The results are shown in Table~\ref{table:coverage_with_bias}.
As expected, the credible regions with the additional bias term have coverage.
\begin{table}
    \centering
    \begin{tabular}{llll} \toprule
        \multicolumn{4}{c}{Empirical coverage probability} \\ \toprule
        {Architecture}\hspace{1cm} & 68\% CR & 95\% CR & 99.7\% CR \\ \midrule
        \multicolumn{4}{l}{$\hat{r}(x\vert \vartheta)$~with~$\vartheta\triangleq(m_\textsc{wdm})$} \\ \midrule
        \textsc{mlp} & $0.704_{~\pm0.004}$\hspace{0.1cm} & $0.972_{~\pm0.002}$\hspace{0.1cm} & $0.999_{~\pm0.000}$  \\
        \textsc{mlp-bn} & $0.706_{~\pm0.003}$ & $0.970_{~\pm0.001}$ & $0.999_{~\pm0.000}$\\
        \textsc{resnet-18} & $0.687_{~\pm0.004}$ & $0.955_{~\pm0.002}$ & $0.998_{~\pm0.000}$  \\
        \textsc{resnet-18-bn} & $0.693_{~\pm0.004}$ & $0.966_{~\pm0.002}$ & $0.999_{~\pm0.000}$  \\
        \textsc{resnet-50} & $0.689_{~\pm0.006}$ & $0.967_{~\pm0.001}$ & $0.998_{~\pm0.000}$ \\
        \textsc{resnet-50-bn} & $0.698_{~\pm0.004}$ & $0.969_{~\pm0.001}$ & $0.999_{~\pm0.000}$ \\ \midrule
        \multicolumn{4}{l}{$\hat{r}(x\vert\vartheta)$~with~$\vartheta\triangleq(m_\textsc{wdm},~$\tage$)$} \\ \midrule
        \textsc{mlp} & $0.704_{~\pm0.004}$ & $0.973_{~\pm0.001}$ & $0.999_{~\pm0.000}$ \\
        \textsc{mlp-bn} & $0.709_{~\pm0.004}$ & $0.970_{~\pm0.001}$ & $0.999_{~\pm0.000}$ \\
        \textsc{resnet-18} & $0.688_{~\pm0.005}$ & $0.965_{~\pm0.002}$ & $0.998_{~\pm0.000}$ \\
        \textsc{resnet-18-bn} & $0.692_{~\pm0.006}$ & $0.967_{~\pm0.002}$ & $0.999_{~\pm0.000}$ \\
        \textsc{resnet-50} & $0.694_{~\pm0.005}$ & $0.968_{~\pm0.002}$ & $0.999_{~\pm0.000}$  \\
        \textsc{resnet-50-bn} & $0.695_{~\pm0.006}$ & $0.968_{~\pm0.001}$ & $0.999_{~\pm0.000}$ \\
        \bottomrule
    \end{tabular}
    \caption{Summary of the coverage diagnostic with an artificially lowered highest density level.
      In doing so, we make the credible intervals more conservative such that the procedure has coverage
      at the specified confidence levels.
      ~~\protect\code{https://github.com/JoeriHermans/constraining-dark-matter-with-stellar-streams-and-ml/blob/master/experiments/experiment-inference/out/summary-coverage-with-cr-bias.ipynb}}
    \label{table:coverage_with_bias}
\end{table}

\section{Comparisons against Approximate Bayesian Computation}
\label{appendix:sec:comparison_abc}
See next page.

\clearpage
\onecolumn

\begin{figure}
    \vspace{1.5cm}
    \centering
    \includegraphics[angle=90,width=.85\linewidth]{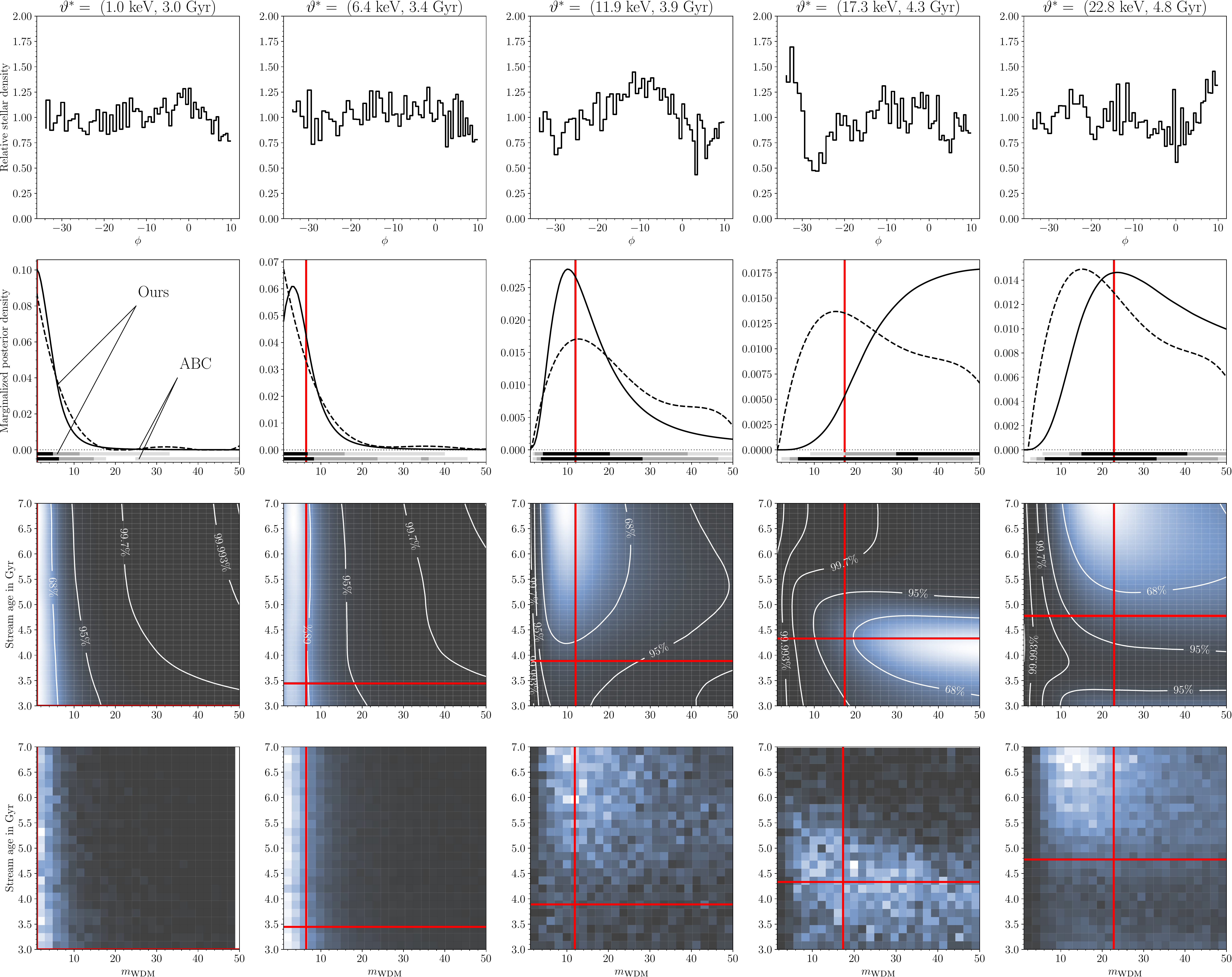}
    \caption{
Direct comparison of ABC against the proposed method. The top row shows the observable. The second row the marginalized posteriors for both methods.
Finally, row 3 and 4 show the joint posterior for our method and ABC respectively.
The nominal target parameter is indicated by the red line.
It is visually apparent that the proposed methodology produces stronger constraints of the groundtruth compared to ABC.
    ~~\protect\code{https://github.com/JoeriHermans/constraining-dark-matter-with-stellar-streams-and-ml/experiments/experiments-inference/out/summary-abc-new.ipynb}}
    \label{fig:abc_new_comparison_panel_1}
\end{figure}
\begin{figure}
    \vspace{4.5cm} 
    \centering
    \includegraphics[trim=0 0 350 0, clip, angle=90,width=.85\linewidth]{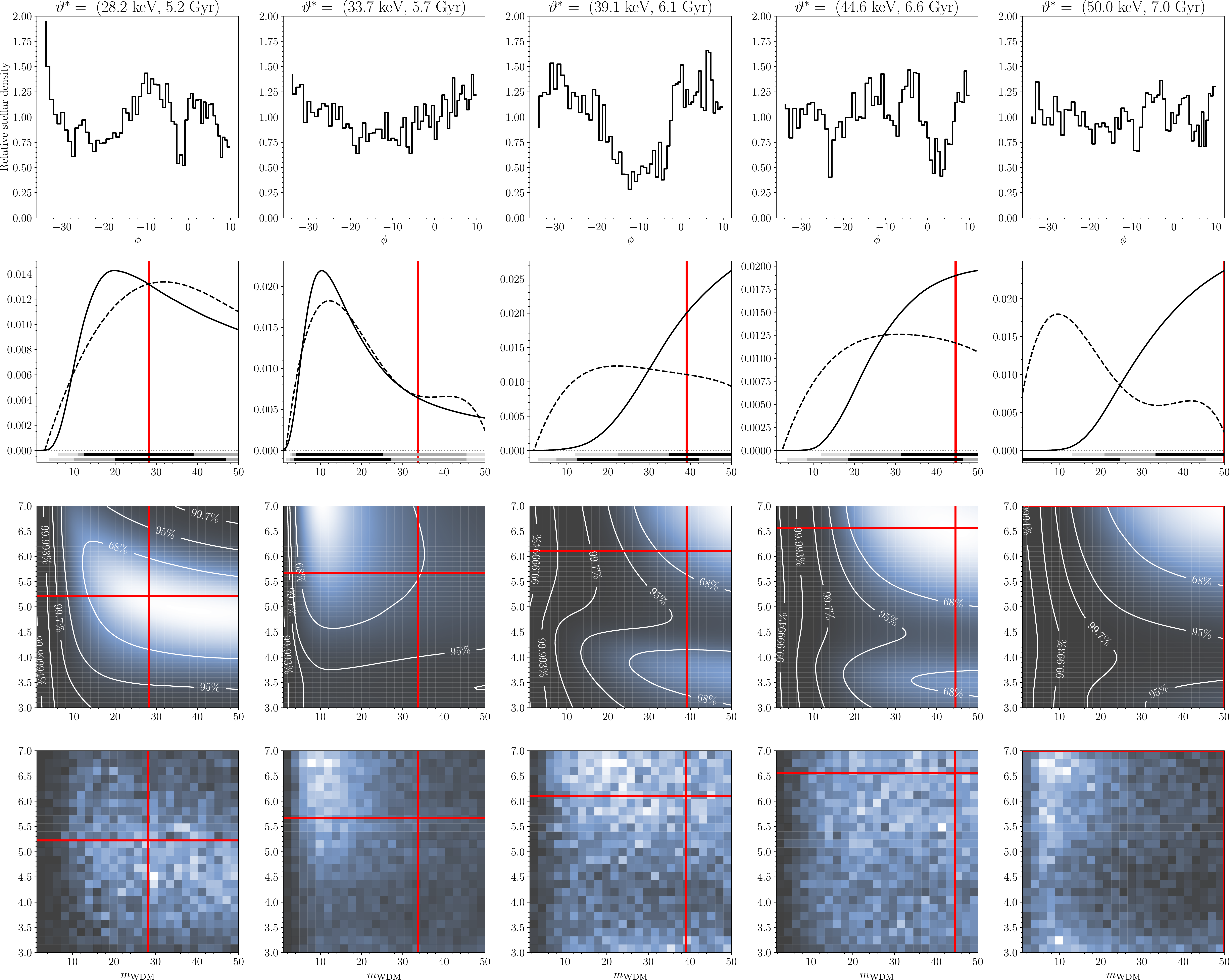}
    \caption{
    Direct comparison of ABC against the proposed method. Refer to Figure~\ref{fig:abc_new_comparison_panel_1} for the initial results.
    ~~\protect\code{https://github.com/JoeriHermans/constraining-dark-matter-with-stellar-streams-and-ml/experiments/experiments-inference/out/summary-abc-new.ipynb} 
    }
    \label{fig:abc_new_comparison_panel_2}
\end{figure}


\clearpage
\bsp	
\label{lastpage}
\end{document}